\documentclass{article}
\usepackage{spconf,amsmath,graphicx}
\usepackage{tikz}
\usepackage{booktabs}
\usepackage{amsmath}
\usepackage{flushend}

\title{A Spatio-Temporal Aligned SUNet Model FOR LOW-LIGHT VIDEO ENHANCEMENT}
\name{Ruirui Lin, Nantheera Anantrasirichai, Alexandra~Malyugina, and David Bull \thanks{This work was supported by UKRI MyWorld Strength in Places Programme (SIPF00006/1)}}
  
  \address{Visual Information Laboratory, University of Bristol, UK}
\begin{document}
%
\maketitle
\begin{abstract}
Distortions caused by low-light conditions are not only visually unpleasant but also degrade the performance of computer vision tasks. The restoration and enhancement have proven to be highly beneficial. However, there are only a limited number of enhancement methods explicitly designed for videos acquired in low-light conditions. We propose a Spatio-Temporal Aligned SUNet (STA-SUNet) model using a Swin Transformer as a backbone to capture low light video features and exploit their spatio-temporal correlations. The STA-SUNet model is trained on a novel, fully registered dataset (BVI), which comprises dynamic scenes captured under varying light conditions. It is further analysed comparatively against various other models over three test datasets. The model demonstrates superior adaptivity across all datasets, obtaining the highest PSNR and SSIM values. It is particularly effective in extreme low-light conditions, yielding fairly good visualisation results. 
\end{abstract}
\section{Introduction}
\label{sec:intro}
Images and videos serve as powerful mediums for capturing moments and conveying information. However, the quality of these visual representations can often be compromised by distortions and noise, often introduced during acquisition. For example, incorrect exposure settings are often the result of an inaccurate balance among the components of the exposure triangle: shutter speed, aperture, and ISO value. This imbalance can lead to a low signal-to-noise ratio (SNR), undesirable noise due to high ISO sensitivity, blurring effects caused by moving objects (low shutter speed), and out-of-focus objects with a shallow depth of field (large aperture). These distortions inevitably lead to poor subjective (perceptual) quality but also degrade the performance of various machine vision tasks such as detection,  classification and tracking as used in surveillance, autonomous driving, medical imaging and natural history filmmaking \cite{Yi:tracking:2023}.  Addressing the challenge of how to reliably and robustly enhance low-light visual content is thus a key component in future video production and analytics pipelines.

\begin{figure}[t]
  \centering
   \includegraphics[width=\linewidth]{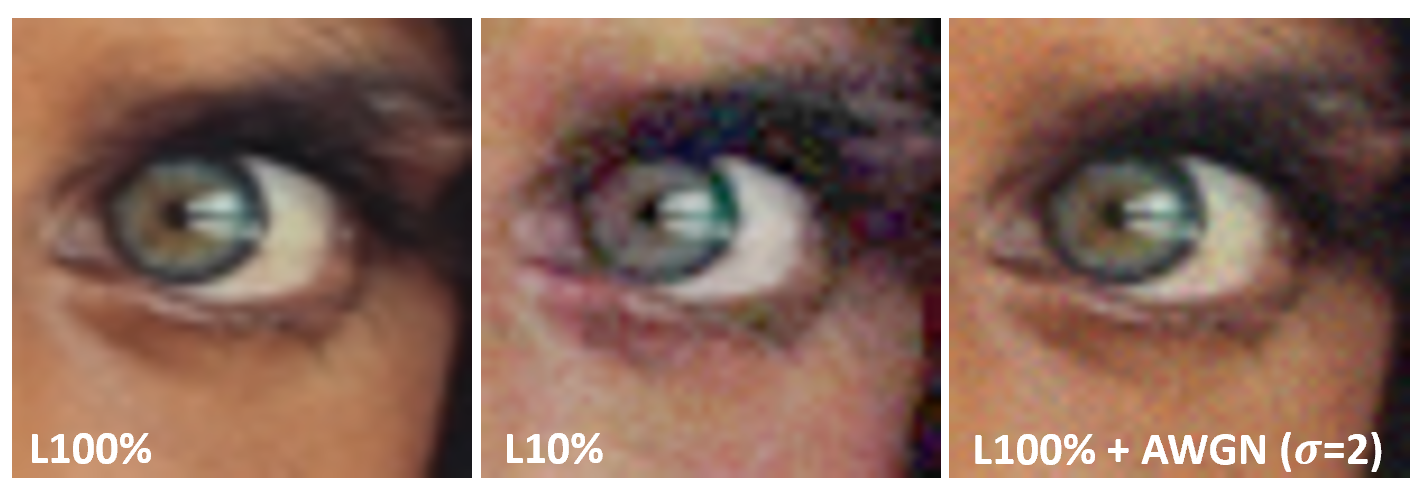} 
  \caption{\small Distortions in cropped images of the `Faces2' sequence. (Left) Normal light. (Middle) Enhanced low light (10\% brightness) using histogram matching to the normal light to visualise distortions under low-light conditions. (Right) Normal light plus Gaussian noise.} 
  \label{fig:noise}
  \vspace{4mm}
   \includegraphics[width=\linewidth]{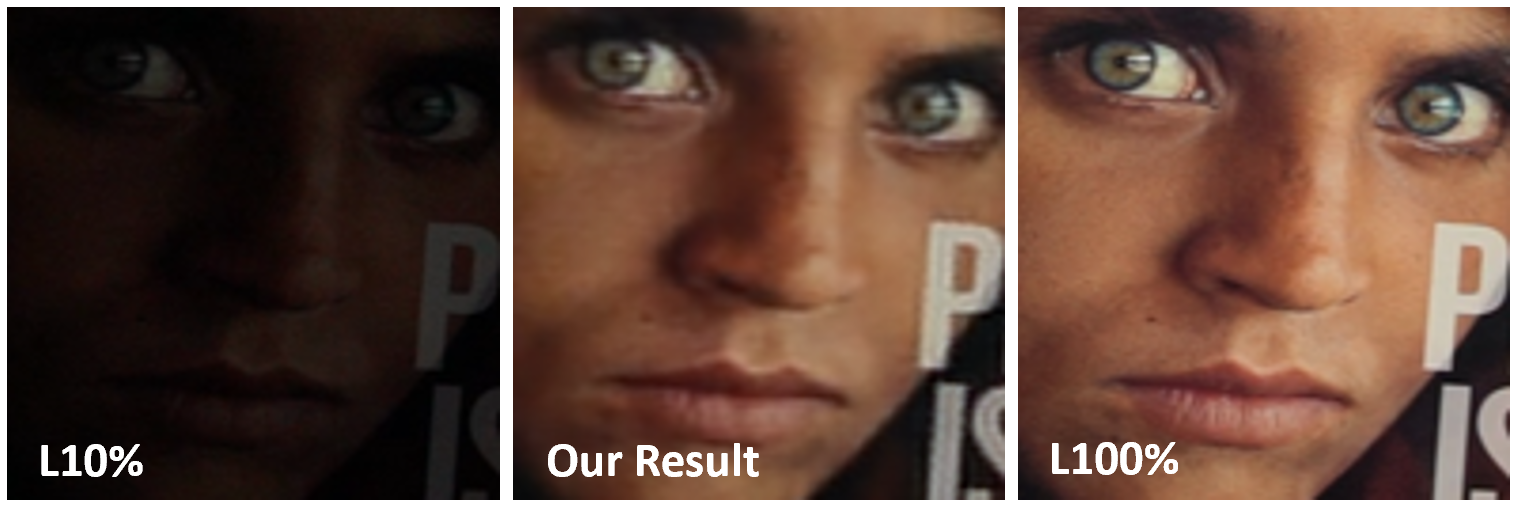} 
  \caption{\small Enhancement visualisation in cropped images of the `Faces2' sequence. (Left) 10\% light level input. (Middle) Enhanced result of our method. (Right) Normal light (100\%).}
  \label{fig:faces2}
\end{figure}

Deep learning techniques have evolved as the most effective tools for image and video processing, in particular the enhancement of low-light content \cite{R2Xu_2022_CVPR,R3Zheng2021AdaptiveUT,R4ma2022fast}. However, the common approach of sequentially applying image-based techniques to each frame in a video often creates temporal inconsistencies and flickering artefacts. This can, in part, be addressed by enhancing quality on a frame-by-frame basis through temporal smoothing \cite{anantrasirichai:Contextual:2021}. Existing methods often use optical flow for alignment but struggle with accurate motion compensation due to occlusion. A further issue that impacts low-light video enhancement is the amount of training data available and its quality and diversity. This is compounded by a lack of accurate ground truth information for imagery acquired in low-light conditions, since gathering clean ground truth with accurate brightness, colour and motion is practically impossible. Without spatio-tempoaral alignment between low-light inputs and their ground truth, it is challenging to accurately evaluate the performance of trained models.

Some methods overcome this limitation by creating synthetic datasets or adopting self-supervised learning methods \cite{R17triantafyllidou2020low}. However, the generalisation on real-world data is often compromised when training on synthetic data since augmenting the dataset with over-simplified synthetic noise, often modelled as zero-mean Gaussian noise, fails to achieve effective results on real world data captured in the wild. This is particularly evident in low-light scenarios where noise characteristics are much more complex than Gaussian noise, as shown in Figure \ref{fig:noise}. This figure shows a cropped book cover with a human eye, taken from a sequence named `Faces2' with fast linear motion in the BVI dataset \cite{datamzny-8c77-23}. When histogram matching is employed on the low light (10\% for noise visualisation), it is noticeable that the shape of the eyeball is slightly shifted and distorted compared to the normal light ground truth. 
This is a typical scenario caused by motion blur. The human eye also experiences significant colour flattening and distortion under low-light conditions. 
Low-light imagery typically loses information during acquisition, including edge distortions, texture loss, motion blur and color changes; these often lead to bias and clipping problems when used to train machine learning methods. Self-supervised learning techniques, although independent of paired input-ground truth datasets and capable of enhancing contrast and brightness by learning subtle content priors, often demand excessive computational complexity. Challenges also arise in the context of color restoration and correction in low-light images where the limited availability of colour information adds to the complexity.

Given that relatively few low-light enhancement methods have been designed specifically for low-light video processing, coupled with the ubiquity and importance of low light content, there is an urgent need for effective low-light video enhancement methods that exploit spatio-temporal correlations. Potential for achieving this is offered by recent advances in vision transformer based mechanisms. Transformers have gained increasing popularity due to their self-attention mechanism, which accelerates pixel-wise optimisation and effectively addresses long-range dependencies in the data \cite{vaswani2023attention}. They have underpinned successes in Natural Language Processing (NLP) and computer vision, outperforming many Convolutional Neural Network (CNN)-based models.

In this paper, we introduce a lightweight STA-SUNet model based on the Swin Transformer \cite{R23Fan_2022} backbone. Traditional convolution, being content-dependent, can raise challenges with larger patches and long-range dependencies, failing to capture global features. SUNet, merging Transformer and UNet, addresses these issues by substituting the convolution layer with the Swin Transformer block, surpassing CNN-based methods in image denoising across well-known benchmark datasets. By realising the importance of frame alignment in video processing, we integrate an additional feature alignment module as proposed in EDVR \cite{R19wang2019edvr}. This module helps to align features spatio-temporally across multiple input frames at the feature level. These aligned features are then passed to SUNet for further feature extraction and reconstruction. An example of our low-light enhancement result is shown in Figure \ref{fig:faces2}.

We train and verify the model with a new high-quality and large input-ground truth aligned dataset. The model achieves effective enhancement of low-light video, addressing many of the previously mentioned challenges. We discuss the implementation, evaluation and analysis of the proposed method, with a comprehensive performance assessment evaluating the effectiveness of the model. 

Our main contributions can be summarised as follows: 
\begin{itemize}
    \item We propose a lightweight Spatio-Temporal Aligned Swin Transformer-based SUNet model (STA-SUNet), specifically designed for low-light video enhancement. Our model provides high adaptivity to natural low-light videos.
    \item We train and test the STA-SUNet model on a novel, high-quality, fully registered dataset \cite{datamzny-8c77-23} captured under different light levels, featuring dynamic scenes. This directly contributes to the increased effectiveness of the enhancement process.
    \item We conduct thorough quantitative and qualitative analyses to evaluate the effectiveness of the proposed method using three different test datasets.
\end{itemize}

\section{Related Work}
\label{sec:work}
Low-light video enhancement is challenging due to the complexity and inconsistency of real low-light conditions, which involve multiple combinations of distortion types. This section discusses prior work on low-light enhancement, including both single-image and video processing. It also describes the datasets utilised in these methods, followed by a brief overview of general video restoration tasks.

\subsection{Low-light image enhancement (LLIE)}
\label{ssec:LLIE}

Before the recent advances in deep learning, traditional digital image processing approaches such as Histogram-equalisation (HE) \cite{R1ibrahim2007brightness}, Retinex theory models \cite{R7fu2015probabilistic}, unsharp masking algorithms \cite{R8unsharp}, BM3D (Block-Matching and 3D filtering) along with its extension to 4D (BM4D)\cite{BM4D}, represented the state of the art. Retinex decomposes an image into reflection and illumination parts using well-designed priors, treating the estimated reflection as the restoration result. Although this contributed to Retinex's popularity, it lacks adaptivity and is associated with lengthy computational time for complex optimisation processes. BM3D and BM4D effectively reduce noise through multidimensional filtering but are less preferred due to the noticeable artefacts they produce. Histogram-equalisation and unsharp masking, while effective, can potentially cause over-enhancement in specific regions, leading to noise amplification. All these methods can result in unrealistic enhancements and undesired artefacts which, in turn, lead to the loss of unique information in areas with distinct details. 

Recent approaches predominantly leverage deep-learning techniques for enhancement. LLNet \cite{R9LORE2017650} initially employed an auto-encoder for low-light image denoising and brightening in 2017. More recently, methods such as addressing low SNR on a spatial basis through an SNR-guided transformer \cite{R2Xu_2022_CVPR}, learning noise through an unfolding variation regularisation model in sRGB space \cite{R3Zheng2021AdaptiveUT},  have also proven to be effective, yielding state-of-the-art results. 

In the context of frame-to-frame processing, single-image processing struggles to aggregate temporal information between frames in video restoration tasks effectively and may face flickering problems if applied to videos directly. This challenge makes it more difficult to adapt single-image enhancement models to video, especially when high levels of motion or dynamic textures are present.

\subsection{Low-light video enhancement (LLVE)}
\label{ssec:LLVE}
Despite the rapid advancements in deep learning for low-light image enhancement, there remains a noticeable gap when it comes to extending these models to enhance low-light video \cite{R24li2021lowlight}. 

Most approaches rely on supervised learning methods, such as ResNet \cite{R13he2015deep} and UNet \cite{R14unet}, for feature extraction. In 2018, Chen et al. introduced the real low-light image SID dataset \cite{R12DBLP:journals/corr/abs-1805-01934}, initially trained using UNet \cite{R14unet}. A Siamese network based on ResNet was proposed in \cite{R159009494} along with the dataset DRV, assuming that modelling with static scenes could generalise well to dynamic scenes. SMOID \cite{R169010274} modified UNet to enable multiple convolutional layers to handle the displacement between frames caused by moving objects. However, its effectiveness is compromised with fast-moving objects exhibiting substantial temporal displacement. 
SDSD \cite{R18Wang2021SeeingDS} introduces two branches of the network for noise reduction and illumination enhancement based on Retinex theory. Issues such as spatial non-alignment \cite{R18Wang2021SeeingDS}, limited motion variation \cite{R159009494}, and data scarcity \cite{R169010274}, further complicate the scenario. Some research methods assume that inputs are in the raw Bayer space \cite{R159009494,R169010274}. However, relying on raw Bayer format in datasets constraints the model's accessibility to the wider research community, as consumer cameras typically do not support raw video formats.

\subsection{Video restoration}
\label{ssec:vr}
As the methods for low-light video enhancement are limited, it is worth considering other video restoration approaches. One such example is EDVR \cite{R19wang2019edvr}, which is implemented with DCN (Deformable Convolution) \cite{R20dai2017deformable} and attention mechanisms to align and fuse features both spatially and temporally. It demonstrates superior performance on video super-resolution and deblurring. Our previous work, integrating feature-level alignment with the traditional UNet (PCDUNet) \cite{datamzny-8c77-23}, reveals effectiveness in enhancing low light videos with motion. Transformer models have found success in natural language processing (NLP) and have been widely applied to vision tasks.
The Swin Transformer \cite{R22liu2021swin} is a state-of-the-art method that addresses issues in pixel-wise vision tasks. It effectively solves problems posed by non-linear, high computational complexity relative to image size, outperforming many convolutional neural networks (CNN)-based methods by constructing hierarchical feature maps with shifted windows. This gave rise to SUNet \cite{R23Fan_2022}, which adopts the Swin Transformer as a backbone and introduces a dual up-sample layer to alleviate the checker-board effect in traditional UNet, achieving excellent results in image denoising.

\section{Proposed Method}
\label{sec:method}
\begin{figure*}[t]
   \includegraphics[width=\linewidth]{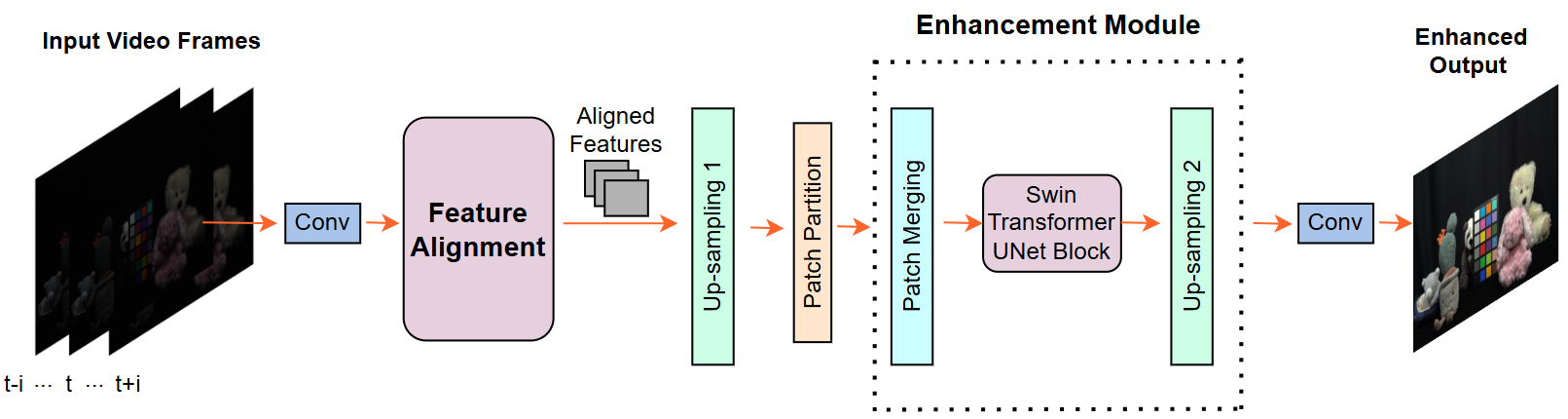}
   \caption{\small Proposed STA-SUNet framework}
   \label{fig:STA-SUNet}
\end{figure*}

The proposed STA-SUNet method, shown in Fig. \ref{fig:STA-SUNet}, is specifically designed for low-light video enhancement. The proposed method starts with aligning multiple input frames at the spatio-temporal feature level. Subsequently, the enhancement is performed using a U-Net-like architecture with Swin Transformer \cite{R22liu2021swin}. The end-to-end framework is trained with L1 loss.  Detailed explanations of the architecture are provided below.



\subsection{Feature alignment}
Multiple input images, denoted by \(I_{t+i},i\in \{-N, -N+1, \ldots , N-1, N \}\), initially proceed to the feature space via convolutional layers before passing through the alignment module. \(N\) represents the number of neighbouring frames at time \(t\). This module aligns features of \(N\) neighbouring frames with the target frame \(I_t\) using a three-level pyramid through deformable convolutions. This aims to utilise information from adjacent frames in a video while minimising disparities. The detailed procedure is described below.

For each layer \(L\), feature maps from two neighboring frames are concatenated and processed through convolutional layers \(L_a\) and \(L_b\) to generate learnable offsets, denoted as\(\delta\). These offsets, combined with adjacent frame features \(I^{L}_{t+i}\), are input into deformable convolution (DC), at that level. The resulting features are then upsampled together with a factor of 2 using bilinear interpolation and fed into the next layer ($L$+1). At each subsequent layer, the offsets generated are concatenated with the upsampled features from the previous layer. The offset-feature prediction process can be represented as follows: 
\begin{equation}
    \begin{gathered}
        \delta^{L}_{t+i} = L_a([I_{t+i},I_t],\delta^{L+1}_{t+i}), \\
        (\hat{I}_{t+i})^L = L_b(DC(I^{L}_{t+i},\delta^{L}_{t+i}),(\hat{I}_{t+i})^{L+1})
    \end{gathered}
\end{equation}

This iterative process spans three layers. A final deformable convolution generates the output of this alignment module, which are aligned features. These aligned features are then upsampled in the first up-sampling layer, as depicted in Figure \ref{fig:STA-SUNet}, using the Pixel Shuffle technique \cite{shi2016realtime}, facilitating patch partitioning in the enhancement module.

\subsection{Enhancement module}
Following Swin Transformer UNet (SUNet) \cite{R23Fan_2022}, the enhancement module extracts and concatenates the features of neighbouring patches through a \(3\times3\) convolution layer and patch merging, which can be viewed as a downsampling process. SUNet replaces traditional UNet convolution layers with Swin Transformer blocks, with 8 layers in each block. These Swin Transformer layers consist of multiple groups of window multi-head self-attention (W-MSA) and shifted-window multi-head self-attention (SW-MSA). To address the lack of connections between windows in the window-based self-attention module, a shifted-window approach is used for cross-window connections among multiple layers, creating a hierarchical representation. An efficient patch computation method known as cycle-shifting is applied to self-attention in the shifted window partitioning, reducing computational time. This approach proves effective when dealing with situations where extra padding would increase computation, especially in the presence of small windows during shifting. The computation of consecutive Swin Transformer blocks in this process follows:   
\begin{equation}
    \begin{gathered}
    \hat{o}^{L} = \text{A}(LN(o^{L-1}))+o^{L-1}, \\
    o^{L} = MLP(LN(\hat{o}^{L})) + \hat{o}^{L}\\
    \hat{o}^{L+1} = \text{B}(LN(o^{L}))+ o^{L}, \\
    o^{L+1} = MLP(LN(\hat{o}^{L+1}))+\hat{o}^{L+1}  
    \end{gathered} 
\end{equation}

Here, A and B represent window-based modules using the normal (W-MSA) and shifted-window (SW-MSA) methods, respectively. The output \(o\) is the output feature for each layer \(L\). A Layer-Norm layer, denoted as \(LN\) is implemented before each MSA module and each MLP (multi-layer perception), with a residual connection applied after each module.
In the second upsampling stage shown in Figure \ref{fig:STA-SUNet}, the traditional transposed convolution is substituted with a combination of Bi-linear and Pixel Shuffle techniques. This blend, referred to as dual upsampling, is implemented to reduce the checker-board effect. Following the original paper, SUNet architecture has 5 layers, utilising skip connections to transfer feature maps from the patch merging stage to the second upsampling stage. Finally, a last \(3\times3\) convolution layer is employed to reconstruct the restored frame with enhanced lightning. 

\section{Experiments}
\label{sec:exp}
\subsection{Dataset}

\begin{figure}[t]
  \centering
   \includegraphics[width=\linewidth]{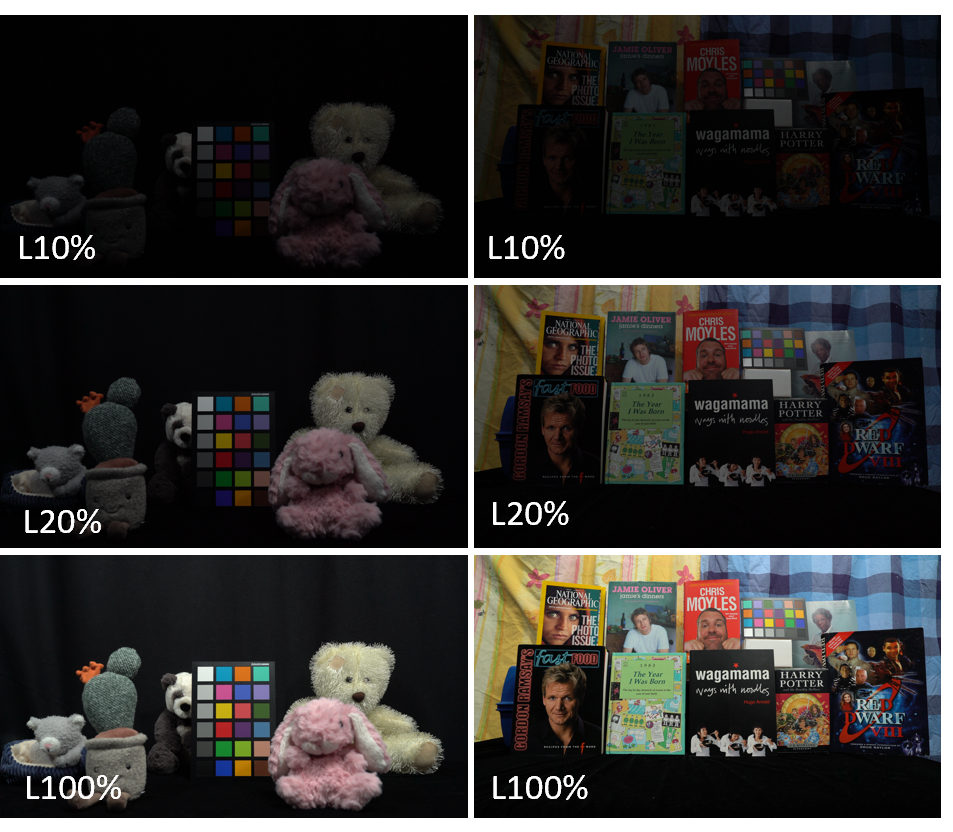} 
  \caption{\small Low light data example: from top to bottom, light levels of 10\%, 20\%, and 100\% (normal light). from left to right, soft toys and books with faces.}
  \label{fig:LL}
\end{figure}

We employed the BVI low-light dataset \cite{datamzny-8c77-23} to evaluate our proposed method. This dataset offers high-quality low-light videos, along with their corresponding normal light ground truths, ensuring full registration in both spatial and temporal dimensions. This supports the training of supervised learning models and facilitates full-reference quality assessment. These videos were captured using a Sony Alpha 7SII camera, with a Kessler CineDrive shuttle dolly system. There are a total of 40 scenes with various contents, textures and motions. As shown in Figure \ref{fig:LL}, each scene provides two low-light levels of 10\% and 20\%, as well as normal lighting (100\% light level). The videos are captured in full HD resolution (1920x1080 pixels) in a standard RGB format. Following \cite{datamzny-8c77-23}, the training and testing sets have been pre-defined, with 32 and 8 scenes used for training and testing, respectively.

\subsection{Experiment settings}
The STA-SUNet network is trained on the BVI dataset without relying on any pre-trained networks. We use 5 RGB frames as inputs, with an image size of 512\(\times\)512, a patch size of 4, a window size of 8, and a batch size of 1. 
The training data is augmented with random flipping and cropping as a pre-processing step. Adam optimiser is used with the initial learning rate of \(1\times10^{-6}\). The loss function used for optimisation is the L1 loss. We implement our model on PyTorch with a single NVIDIA GeForce RTX 3090 GPU. The metrics used for evaluation are PSNR (Peak Signal-to-Noise Ratio) and SSIM (Structural Similarity Index Measure), where higher values indicate better performance. 

\subsection{Impact of different light levels}

The impact of light levels on the model's performance is analysed and presented below through quantitative analysis and illustrative figures. The STA-SUNet model is firstly trained on a dataset at light levels of 10\%, and 20\%, respectively, and then on the collective dataset of light levels at both 10\% and 20\%. Table \ref{tab:studylightlevels} summarises the results. It is observed that the model, when trained and tested using data at the same light level, achieves better performance than when using data with different light levels. However, the model trained under extremely low lighting conditions, such as 10\% light level only, tends to perform poorly when tested with data in brighter light conditions, such as 20\% light level. This is indicated by PSNR and SSIM values of 11.30 and 0.633, respectively, in the first row on Table \ref{tab:studylightlevels}. Conversely, the model trained with data under 20\% light levels only tends to perform worse when tested with data taken from 10\% light conditions, as shown by PSNR and SSIM values in the second row of Table \ref{tab:studylightlevels}. The model demonstrates better performance when trained across a diverse dataset at different light levels, as illustrated in the third column in Table \ref{tab:studylightlevels}. 

\begin{table}[t]
    \centering
     \caption{\small Performance of the STA-SUNet model when trained and tested with different light levels}
     \small
    \begin{tabular}{c|ccc}
    \toprule
    test & 10\% & 20\% & 10\%+20\%  \\ \hline
        train & PSNR/SSIM & PSNR/SSIM & PSNR/SSIM  \\ \hline
        10\% & 27.32 / 0.847& 11.30 / 0.633 & 19.32 / 0.740\\
        20\% & 12.04 / 0.630 & 31.41 / 0.930 & 21.74 / 0.781\\
        10\%+20\% & 24.44 / 0.822 & 27.84 / 0.876 & 26.14 / 0.849\\
    \bottomrule
    \end{tabular}
    \label{tab:studylightlevels}
\end{table} 

\begin{figure*}[t]
   \includegraphics[width=\linewidth]{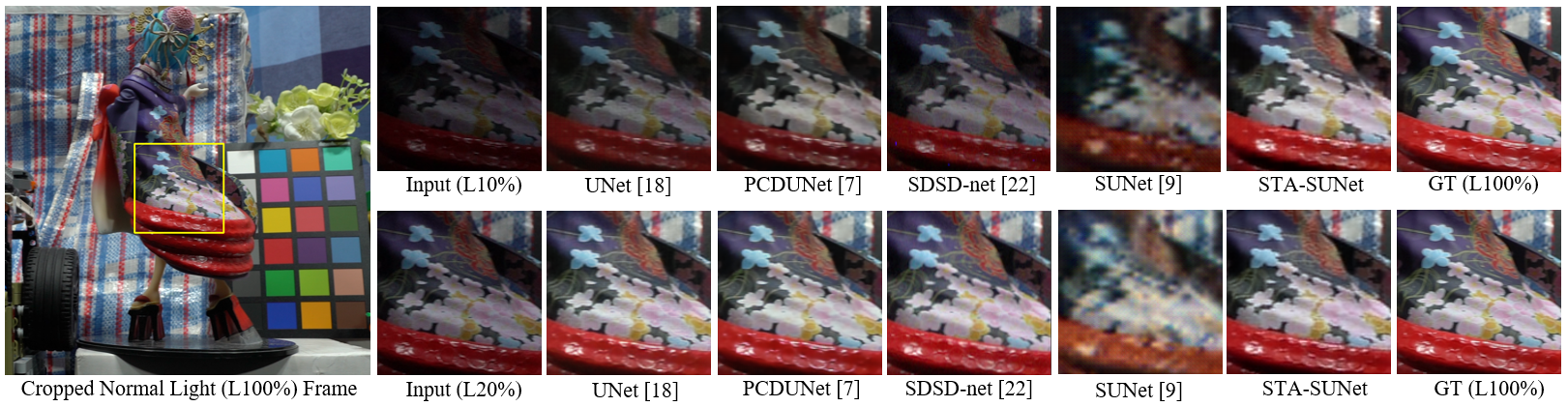}
   \caption{\small Visualisation results when using 5-frame inputs comparison for low-light enhancement on cropped images of frame 102 in the `Figures2' sequence from the BVI dataset. (Top) From left to right, 10\% light level input, enhanced results, and 100\% normal light groundtruth. (Bottom) From left to right, 20\% light level input, enhanced results, and 100\% normal light groundtruth.}
   \label{fig:output}
\end{figure*}

We further investigated how our proposed method deals with low-light distortion. First, we mitigated the influence of varying brightness and color balance as follows. Before feeding the test data into the model, histogram matching is conducted to align the histogram of the test data with that of the training data, leaving noise in the results. The PSNR and SSIM values demonstrate a significant improvement upon such post-processing, as depicted in Table \ref{tab:postprocesslightlevels}. This suggests that restoring brightness and colour changes is a challenging task. Fine-tuning state-of-the-art denoisers may not yield immediate results.


\begin{table}[t]
    \centering
     \caption{\small Performance of the STA-SUNet model when trained with light levels of 10\% and 20\% and tested with histogram matching}
     \small
    \begin{tabular}{c|ccc}
    \toprule
    \multicolumn{4}{c}{With histogram matching} \\ \hline
    test & 10\% & 20\% & 10\%+20\% \\ \hline
        train & PSNR/SSIM & PSNR/SSIM & PSNR/SSIM \\ \hline
        10\% & 27.32 / 0.847 & 27.28 / 0.860 & 27.30 / 0.853 \\
        20\% & 27.15 / 0.811 & 31.41 / 0.930 & 29.28 / 0.871\\
    \bottomrule
    \end{tabular}
    \label{tab:postprocesslightlevels}
\end{table}

\subsection{Impact of number of input frames}
Table \ref{tab:frame} shows the test results for different numbers of input frames trained on the mixed light samples and tested on the 10\% light level. Through our experiments we observe that, in general, an increasing number of input frames from 1 to 5 improves the resulting enhanced quality as temporal consistency improves. While using only three input frames proves insufficient for gathering enough temporal information. The advantage of using multiple frames as input is particularly relevant in processing video sequences, where spatio-temporal correlations often exist among adjacent frames. This obviously trades off against memory requirements.

\begin{table}[h]
    \centering
    \caption{\small Performance of the STA-SUNet model when trained on different numbers of input frames}
    \small
    \begin{tabular}{c|ccc}
    \toprule

        Input frame & 1 & 3 & 5  \\ \hline
        PSNR/SSIM & 18.41 / 0.677 & 18.45 / 0.680 & 24.44 / 0.822 \\
    \bottomrule
    \end{tabular}
    \label{tab:frame}
\end{table}

\subsection{Performance comparison}
\begin{table}[t]
    \centering
    \caption{\small Performance comparison of the models tested on different datasets (trained on the BVI dataset).}
    \small
    \begin{tabular}{c|ccc}
    \toprule
        Dataset & DRV \cite{R159009494} & SDSD \cite{R18Wang2021SeeingDS}  & BVI \cite{datamzny-8c77-23}  \\ \hline
        Methods & PSNR/SSIM & PSNR/SSIM & PSNR/SSIM \\ \hline
        UNet \cite{R14unet}  & 11.39 / 0.282 & 16.89 / 0.667 & 21.82 / 0.777 \\
        PCDUNet \cite{datamzny-8c77-23} & 18.52 / 0.696 & 19.36 / \textbf{0.729}  & 24.51 / 0.845\\
        SDSD-net \cite{R18Wang2021SeeingDS} & 16.18 / 0.595 & 18.83 / 0.688 & 10.30 / 0.476 \\ 
        SUNet \cite{R23Fan_2022} & 17.70 / 0.599 & 14.89 / 0.635 & 18.95 / 0.645 \\ \hline
        \textbf{STA-SUNet} & \textbf{18.74} / \textbf{0.699} & \textbf{19.97} / 0.714 & \textbf{26.14} / \textbf{0.873}\\  
    \bottomrule
    \end{tabular}
    \label{tab:alltest}
\end{table}

\begin{table}[t]
    \centering
    \caption{\small Average performance comparison}
    \small
    \begin{tabular}{c|c}
    \toprule
        Methods &  Average PSNR/SSIM \\ \hline
        UNet \cite{R14unet}  & 16.70 / 0.575 \\
        PCDUNet \cite{datamzny-8c77-23} & 20.79 / 0.757 \\
        SDSD-net \cite{R18Wang2021SeeingDS} & 15.10 / 0.586 \\ 
        SUNet \cite{R23Fan_2022} & 17.18 / 0.626\\ \hline
        \textbf{STA-SUNet} & \textbf{21.62} / \textbf{0.762} \\  
    \bottomrule
    \end{tabular}
    \label{tab:avgtest}
\end{table}

\begin{figure}[h]
  \centering
   \includegraphics[width=\linewidth]{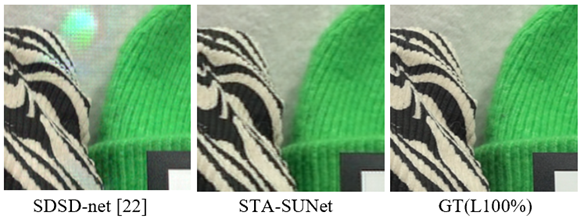} 
  \caption{\small Visualisation comparison between STA-SUNet and SDSD-net: from left to right, SDSD-net result, STA-SUNet result and normal light 100\%.}
  \label{fig:hats}
\end{figure}

To further explore the effectiveness of the STA-SUNet model, we conduct a comparative performance evaluation by testing the model on three datasets, namely, DRV \cite{R159009494}, SDSD \cite{R18Wang2021SeeingDS}, and BVI \cite{datamzny-8c77-23}. The DRV data undergoes an adjustment process, with its green channel reduced by half to standard RGB, ensuring a fair comparison. The objective results are illustrated in Table \ref{tab:alltest}. The last column highlights the importance of training on a fully-registered dataset, with the STA-SUNet model achieving the highest PSNR values. Table \ref{tab:avgtest} presents the average objective results across all three test datasets. In general, the STA-SUNet model demonstrates superior adaptivity across all datasets, achieving the highest average PSNR and SSIM values. The visualisation results are shown in Figure \ref{fig:output} with cropped images of frame 102 in the `Figures2' sequence. It is evident from the results that the STA-SUNet model demonstrates outperforms, especially under extreme low-light conditions (10\% light level), producing good visual results. While 20\% light levels are generally easier for enhancement models, both PCDUNet and UNet exhibit over-exposure at this level, even though PCDUNet has similar objective performance to STA-SUNet. Undesired speckle effect in the output of SDSD-net, as shown in Figure \ref{fig:hats}, contributes to its low objective results. 

\section{Conclusion}
\label{sec:conclude}
We propose a Spatio-Temporal Aligned SUNet (STA-SUNet) model, leveraging the Swin Transformer backbone, to enhance low-light videos. Trained and validated on a novel, fully-registered dataset with diverse motions, our model's effectiveness is evaluated on two other datasets as well. Through extensive experiments, we analyze the impact of various light levels and the temporal consistency. Quantitative results show the STA-SUNet outperforming other models in terms of PSNR and/or SSIM across all datasets. Comparative analyses against UNet, PCDUNet, SDSD-net, and SUNet are conducted. By achieving the highest average PSNR and SSIM values across all datasets, our model demonstrates superior adaptivity, especially in extreme low-light conditions, yielding fairly good visualisation results.

\vfill\pagebreak


\bibliographystyle{IEEEbib}
\bibliography{refs}

\end{document}